\documentclass[aps,prl,twocolumn,groupedaddress]{revtex4}

\usepackage{graphicx}
\usepackage{color}

\begin{document}

\title{Reply to Comment on ``Why is the DNA Transition First Order?'' and
``Griffiths Singularities in Unbinding of Strongly Disordered
Polymers'' by M. Ya. Azbel}

\date{\today}

{\bf Kafri, Mukamel and Peliti Reply}: In the opening statement of
his Comment \cite{Azb}, Azbel summarizes the main results of the
two papers which are the subject of his Comment by stating that
"They find a first order phase transition \cite{pap2} when
disorder is strong (i.e. the ratio $v$ of the binding energies is
large; in DNA $v$ is ~1.1) and the Griffiths singularity, i.e.
infinite order transition, otherwise \cite{pap1}". In fact, the
two papers claim quite the opposite. In \cite{pap2} it is found
that when disorder is large, the system exhibits a very weak,
infinite order, Griffith type singularity at a low temperature.
This is {\it in addition} to the melting transition which takes
place at a higher temperature, and on which nothing is said in
\cite{pap2}. In \cite{pap1} it is demonstrated that in the
homogeneous case a first order melting transition takes place.

Let us briefly respond to the specific claims on each of the
papers. Addressing the paper ``Griffiths Singularities in
Unbinding of Strongly Disordered Polymers'' \cite{pap2}, Azbel
first claims that there is no Griffiths type singularity in the
{\it unbinding transition}. We stress that our model is not aimed
at analyzing the unbinding transition, as is stated very clearly
in our paper. Our model comprises of two polymers bound by either
weak bonds (with an energy of $O(1)$) or strong bonds (with an
energy of $O(v) \gg 1$). It is found that the model exhibits a
Griffiths type singularity at a low temperature $T_G = O(1)$ in
addition to the melting transition which takes place at a much
higher temperature $T_M=O(v)$. The low temperature transition
corresponds to the unbinding of polymer stretches with a low
binding energy of $O(1)$. The model is analyzed exactly in the
limit where the binding energy $v=\infty$. Since the low
temperature transition does not involve the unbinding of the
strong bonds, the nature of the transition is expected to remain
unchanged when the binding energy $v$ is finite but large. The
high temperature transition has recently been analyzed in
\cite{Tang} and \cite{Azbel} with conflicting conclusions.

Next, Azbel claims that the proof of the existence of Griffiths
type singularity presented in \cite{pap2} is invalid. This
statement is not substantiated by any argument and is simply
wrong. In the paper we have demonstrated that the free energy is
singular at $T_G$, by showing that the zeros of the partition sum
accumulate arbitrarily close to the real temperature axis in the
thermodynamic limit. In addition it is shown that all derivatives
of the free energy are finite at this temperature. The singularity
is thus by definition of a Griffiths type. Azbel also erroneously
state that the free energy Eq. (17) yields a {\it second order}
phase transition while it is evident that all derivatives of the
free energy are finite at the transition temperature, making the
transition infinite order.

In his comment on the second paper ``Why is the DNA Transition
First Order?'' \cite{pap1}, Azbel misinterprets the paper and
applies its conclusions to the unbinding transition of an
infinitely long heteropolymer. In our paper we analyzed the
unbinding of a {\it homopolymer} and showed that when excluded
volume effects are properly taken into account the melting
transition is first order. This result may be used to understand
the experimental observation that in DNA molecules which are a few
thousands base pairs long the unbinding takes place by a series of
steps which are sharp. Each step corresponds to the unbinding of a
finite region of  a few hundred base pairs long within the DNA
molecule. The fact that the steps are sharp but somewhat rounded
may be understood as a broadening of the first order transition
resulting from the finite length of the region and the
heterogeneity of the molecule. Clearly when one considers the
melting of a very long DNA molecule (of the order of $10^6$ base
pairs) the steps are averaged out and the resulting melting
transition is of a different nature which has to be properly
analyzed. One should not use our results, obtained for the melting
of a homopolymer, to analyze the melting of a very long
heteropolymer, as was done by Azbel in his comments. Our results
are applicable for the understanding of the sharp steps which take
place in the melting curve of not too long polymers of the order
of a few thousands base pairs long.

Y. Kafri\\
Department of Physics, Harvard University, Cambridge, MA 02138.\\

D. Mukamel\\
Department of Physics of Complex Systems, Weizmann Institute of
Science, Rehovot, Israel 76100.\\

L. Peliti\\
Dipartimento di Scienze Fisiche and Unit\`{a} INFM, Universit\`{a}
``Federico II'', Complesso Monte S. Angelo, 80126 Napoli, Italy.

\end{document}